# ON QUANTUM-GEOMETRIC CONNECTIONS AND PROPAGATORS IN CURVED SPACETIME[*]

**Eduard Prugovecki**

*Department of Mathematics*
*University of Toronto*
*Toronto, Canada M5S 1A1*

**Abstract.** The basic properties of Poincaré gauge invariant Hilbert bundles over Lorentzian manifolds are derived. Quantum connections are introduced in such bundles, which govern a parallel transport that is shown to satisfy the strong equivalence principle in the quantum regime. Path-integral expressions are presented for boson propagators in Hilbert bundles over globally hyperbolic curved spacetimes. Their Poincaré gauge covariance is proven, and their special relativistic limit is examined. A method for explicitly computing such propagators is presented for the case of cosmological models with Robertson-Walker metric.

PACS numbers: 0420, 0240, 0220

---

[*] Supported in part by NSERC research grant no. A5206

# 1. Introduction

A recent path-integral formulation [1,2] of quantum propagation in curved spacetime is based on the extrapolation of the strong equivalence principle from the classical to the quantum general relativistic regime. Foundational considerations [3,4] show that Planck's length $\ell_P$ imposes an upper bound on the precision of quantum particle and field localizability, so that it presents an impediment to this extrapolation within the conventional framework for relativistic quantum theory. Therefore, such a formulation of quantum propagation was achieved within a quantum-geometric framework incorporating a fundamental length $\ell \geq \ell_P$, for which, on measurement theoretical grounds [3,4], the most natural candidate is the Planck length $\ell_P$ itself.

In this framework quantum-geometric propagation can be formulated for any curved spacetime represented by a globally hyperbolic Lorentzian manifold $(\mathbf{M},\mathbf{g})$. Its central idea consists of replacing the broken polygonal paths originally introduced by Feynman [5,6] in flat spacetimes with the arcs of geodesics of the Levi-Civita connection determined by $\mathbf{g}$. In the resulting "sum-over paths" for quantum-geometric propagators the free-fall propagation along each such broken geodesic path takes place by the parallel transport governed by a quantum connection obtained by extending the Levi-Civita connection from the Lorentz frame bundle to the Poincaré frame bundle over $(\mathbf{M},\mathbf{g})$. This quantum connection gives rise to infinitesimal parallel transport that coincides with special relativistic quantum evolution. In this manner the strong equivalence principle is extrapolated from the classical to the quantum general relativistic regime.

For the sake of simplicity, in the present note we shall concentrate on the quantum-geometric propagation of single massive bosons of zero spin.

In Secs. 2 and 3 we outline the construction of a Hilbert bundle of geometrically local [2] quantum states for such bosons, and of the corresponding quantum connection governing their quantum-geometric propagation. In Sec. 4 we prove the Poincaré gauge covariance of boson quantum-geometric propagators corresponding to geometrodynamic evolutions [7] that give rise to globally hyperbolic curved spacetimes. In Sec. 5 we show how those propagators can be computed in the case of Roberston-Walker spatially flat spacetimes. This outcome graphically illustrates how quantum-geometric propagation merges into conventional special relativistic propagation in the case where the cosmological expansion factor [7] $a(t)$ in the Roberston-Walker metric is actually a constant, so that the resulting Roberston-Walker spacetime can be identified with Minkowski space.

## 2. Typical fibres for massive bosons of zero spin

For any choice $\ell > 0$ of fundamental constant in Planck natural units, the typical fibre $\mathbf{F}$ of a Klein-Gordon quantum bundle for spin-0 massive bosons adopted in [1,2] consists of all the wave functions

$$\varphi(\zeta) = \tilde{Z}_{\ell,m}^{-1/2} \int_{k^0 > 0} \exp(-i\,\zeta \cdot k)\, \tilde{\varphi}(k)\, d\Omega_m(k) \; , \tag{2.1a}$$

$$\zeta = q - i\ell v \; , \qquad d\Omega_m(k) = \delta(k^2 - m^2)\, d^4k \; , \tag{2.1b}$$

$$q \in M^4 = (\mathbf{R}^4, \eta) \; , \qquad v \in V^+ = \left\{ u \,\middle|\, u^2 = \eta_{\mu\nu} u^\mu u^\nu = 1 \; , \; u^0 > 0 \right\} \; , \tag{2.1c}$$

obtained as $\tilde{\varphi}$ varies over all wave functions in the momentum representation. In (2.1c) $\eta$ denotes the Minkowski metric with the choice +1,–1,–1,–1 for its diagonal elements. The values of the 4-tuples $q$ vary over the 4-dimensional Minkowski space $M^4$ with this metric, and those of $v$ over the corresponding 4-velocity hyperboloid $V^+$. Hence, the wave functions in (2.1a) can be viewed as being defined over the special relativistic phase space $M^4 \times V^+$.

It turns out [8,9,10] that the family of such wave functions constitutes a Hilbert space with the inner product

$$\left\langle \varphi_1 \middle| \varphi_2 \right\rangle = \int \varphi_1^*(\zeta) \varphi_2(\zeta)\, d\Sigma(\zeta) \; , \tag{2.2}$$

where the integration can be carried out with respect to the (unique modulo a multiplicative constant [11]) invariant phase space measure

$$d\Sigma(\zeta) = 2 v^\mu d\sigma_\mu(q) d\Omega(v) \; , \qquad d\Omega(v) = \delta(v^2 - 1)\, d^4v \; , \tag{2.3}$$

over any hypersurface $\sigma \times V^+$ in $M^4 \times V^+$ for which $\sigma$ is a spacelike Cauchy hypersurface in $M^4$. The value of the normalization constant in (2.1a) can be expressed in terms of modified

Bessel function $K_2$ (cf. [2], Secs. 3.7, 3.8 and 4.1)

$$\tilde{Z}_{\ell,m} = 8\pi^4 K_2(2\ell m)/\ell m^2 \ , \tag{2.4}$$

and it is uniquely determined by the requirement that the inner product in in the Hilbert space $L^2(V_m^+)$ be equal to that in (2.2), so that

$$\langle \varphi_1 | \varphi_2 \rangle = \int_{k^o > 0} \tilde{\varphi}_1^*(k) \tilde{\varphi}_2(k) \, d\Omega_m(k) \ , \tag{2.5}$$

and therefore so that the map $\tilde{\varphi} \mapsto \varphi$ determined by (2.1a) would be unitary.

Originally, such Hilbert spaces as **F**, which carry inner products of the type (2.2), were introduced [8,9] in order to resolve foundational problems [12] related to the nonexistence [13] of Poincaré gauge invariant and locally conserved probability currents for positive-energy solution of the Klein-Gordon equation in configuration space. That problem was resolved in the context of such spaces by proving [8,10] that the probability current

$$j^\mu(q) = 2 \int_{V_m^+} v^\mu \left| \varphi(q - i\ell v) \right|^2 d\Omega(v) \ , \tag{2.6}$$

with positive-definite timelike component, is conserved and special-relativistically covariant

$$\partial_\mu j^\mu(q) = 0 \ , \qquad \partial_\mu = \partial/\partial q^\mu \ , \tag{2.7a}$$

$$j^\mu(q) \mapsto j^\mu(q') = \Lambda^\mu{}_\nu j^\nu(q) \ , \qquad q' = a + \Lambda q \ , \tag{2.7b}$$

under the Poincaré transformations which are given by the following unitary representation of the (restricted) Poincaré group $ISO_0(3,1)$:

$$U(a,\Lambda) : f(\zeta) \mapsto f'(\zeta) = f(\Lambda^{-1}(q-b) - i\ell\Lambda^{-1}v) \ , \quad (a,\Lambda) \in ISO_0(3,1) \ . \tag{2.8}$$

This enabled a consistent physical interpretation of the resulting (stochastic) phase space wave functions.

For example, in the context of measurement-theoretical schemes [14,15] capable of dealing with simultaneous measurements of (non-sharp) stochastic position and momentum,

$$P_\varphi(B) = \int_B |\varphi(\zeta)|^2 d\Sigma(\zeta) \ , \qquad \|\varphi\| = 1, \quad \varphi \in \mathbf{F} \ , \tag{2.9}$$

was interpreted [9,10] as being the probability of detection within a (Borel) subset $B$ in any of the aforementioned hypersurfaces $\sigma \times V^+$ of a boson prepared in the state represented by the state vector $\varphi \in \mathbf{F}$. In the nonrelativistic limit, obtained by using generic units and then letting $c \to +\infty$, the agreement of this interpretation with Born's interpretation in orthodox quantum mechanics was then established upon letting $\ell \to +0$ (cf. [10], Secs. 1.6 and 2.6).

On the other hand, if the sharp-point limit $\ell \to +0$ is taken first, then, as can be seen from (2.4), divergences manifest themselves:

$$\tilde{Z}_{\ell,m} = (4\pi^4/\ell^3 m^4) + O(\ell^{-1}) \xrightarrow[\ell \to +0]{} +\infty \ . \tag{2.10}$$

These divergences persist even after the transition to the following alternative form [8,10],

$$\langle \varphi_1 | \varphi_2 \rangle = i\hat{Z}_{\ell,m} \int_{v^0 > 0} \varphi_1^*(\zeta) \overleftrightarrow{\partial}_\mu \varphi_2(\zeta) \, d\sigma^\mu(q) d\Omega(v) \ , \tag{2.11a}$$

of the inner product in (2.2) is made, since then we have [2,10]

$$\hat{Z}_{\ell,m} = \mathrm{K}_2(2\ell m)/m\,\mathrm{K}_1(2\ell m) = (1/\ell m^2) + O(1) \xrightarrow[\ell \to +0]{} +\infty \ , \tag{2.11b}$$

where $\mathrm{K}_1$ and $\mathrm{K}_2$ denote modified Bessel functions.

A close connection with conventional relativistic quantum theory can be nevertheless established, due to the fact that, on one hand

$$K^{(\ell,m)}(q'' - i\ell v''; q' - i\ell v') = \tilde{Z}_{\ell,m}^{-1} \int_{V_m^+} \exp\{[i(q'-q'') - \ell(v'+v'')] \cdot k\} \, d\Omega_m(k) \ , \tag{2.12}$$

is a reproducing kernel, so that for any $\varphi \in \mathbf{F}$,

$$\varphi(\zeta) = \int K^{(\ell,m)}(\zeta; \zeta') \varphi(\zeta') d\Sigma(\zeta') \ , \tag{2.13}$$

and it has the following properties

$$K^{(\ell,m)}(\zeta''; \zeta') = K^{(\ell,m)*}(\zeta'; \zeta'') = \int K^{(\ell,m)}(\zeta''; \zeta) K^{(\ell,m)}(\zeta; \zeta') d\Sigma(\zeta) \ , \tag{2.14}$$

characterizing propagators; whereas, on the other hand, in the sharp-point limit $\ell \to +0$, the geometro-stochastic propagator in (2.12) converges upon renormalization,

$$i(2\pi)^{-3} \tilde{Z}_{\ell,m} K^{(\ell,m)}(\zeta''; \zeta') \xrightarrow[\ell \to +0]{} K_F(q''-q') \ , \qquad q''^0 > q'^0 \ , \tag{2.15}$$

and in the sense of distributions, to the following Feynman propagator $K_F$:

$$K_F(x) = -\frac{1}{(2\pi)^4} \lim_{\varepsilon \to +0} \int d^4k \, \frac{\exp(-ik \cdot x)}{k^2 - m^2 + i\varepsilon} = \frac{i}{(2\pi)^3} \theta(x^0) \int_{V_m^+} \exp(ik \cdot x) \, d\Omega_m(k). \tag{2.16}$$

As shown in [1,2], these facts provide a solution to the problem of divergencies in rela-tivistic quantum field theory, since the Feynman propagator is a singular distribution rather than an ordinary function; whereas the geometro-stochastic propagator in (2.12) is a smooth function, which assumes in its domain of definition the following finite values [2,10]:

$$K^{(\ell,m)}(\zeta''; \zeta') = 2\pi m \tilde{Z}_{\ell,m}^{-1} \, \mathrm{K}_1\!\left(m\sqrt{-\zeta_\Delta \cdot \zeta_\Delta}\right)\!\Big/\!\sqrt{-\zeta_\Delta \cdot \zeta_\Delta} \ , \tag{2.17a}$$

$$\zeta_\Delta^\mu = (q'^\mu - q''^\mu) + i\ell(v'^\mu + v''^\mu) \ , \quad \zeta_\Delta \cdot \zeta_\Delta = \eta_{\mu\nu} \zeta_\Delta^\mu \zeta_\Delta^\nu \ . \tag{2.17b}$$

Due to its exclusive dependence on the Minkowski inner product in (2.17b), the special relativistic invariance of the geometro-stochastic propagator in (2.17a) is manifest.

## 3. Hilbert bundles over curved spacetime

Let us denote by $P\mathbf{M}$ the Poincaré frame bundle over a Lorentzian spacetime manifold $(\mathbf{M}, g)$. This principal bundle consists of all the Poincaré frames $\mathbf{u} = \{(\mathbf{a}, \mathbf{e}_i) | i = 0, 1, 2, 3\}$ above all base locations $x \in \mathbf{M}$, obtained by translating in the tangent space $T_x \mathbf{M}$ in the amount represented by the 4-vector $\mathbf{a}$ a future-oriented and spatially right-handed local Lorentz frame $\{\mathbf{e}_i | i = 0, 1, 2, 3\}$ with origin at the point of contact of the tangent space $T_x \mathbf{M}$ and $\mathbf{M}$. Such a the Poincaré frame bundle has the Poincaré group $\mathrm{ISO}_0(3,1)$ as its structure group [16-18].

The Hilbert bundle $\mathbf{E}$ of geometrically local quantum states for massive spin-0 bosons is defined by $\mathbf{G}$-product of $P\mathbf{M}$ with the typical fibre $\mathbf{F}$ described in Sec. 2,

$$\mathbf{E} = P\mathbf{M} \times_\mathbf{G} \mathbf{F} \ , \qquad \mathbf{G} = \mathrm{ISO}_0(3,1). \tag{3.1}$$

Hence, by the definition of $\mathbf{G}$-products (cf., e.g., [2], Sec. 4.1; [17], Sec. 3.3), the generic element $\Psi$ of the quantum above a given base location $x \in \mathbf{M}$ is an equivalence class

$$\Psi = \{(\mathbf{u}, \Psi) \cdot g^{-1} \, | \, g = (a, \Lambda) \in \mathrm{ISO}_0(3,1)\} \ , \quad (\mathbf{u}, \Psi) \in P\mathbf{M} \times \mathbf{F} \ . \tag{3.2}$$

Such an equivalence class results from the action from the right [16,17] $g : \mathbf{u} \to \mathbf{u} \cdot g$ of the elements $g = (a, \Lambda)$ of the structure group $\mathbf{G} = \mathrm{ISO}_0(3,1)$ upon the Poincaré frames $\mathbf{u}$ within the fibre of $P\mathbf{M}$ above $x$, and from the action from the left of the representation in (2.8) upon the elements $\Psi$ of the typical fibre $\mathbf{F}$, so that, by definition[1]

---

[1] The use of the inverse of $g$ in (3.2) and (3.3) secures the agreement of some later formulae with the conventions in physics literature and with Eq. (1.11) in Chapter 4 of [2], in which $\mathbf{u} \cdot (b, \Lambda)$ should be $\mathbf{u} \cdot (b, \Lambda)^{-1}$.

$$(\boldsymbol{u}, \Psi) \cdot g^{-1} = (\boldsymbol{u} \cdot g^{-1}, U(g)\Psi) \;, \quad g = (a, \Lambda) \in \mathrm{ISO}_0(3,1) \;. \tag{3.3}$$

This construction gives rise to the diffeomorphisms

$$\boldsymbol{\sigma}_x^{\boldsymbol{u}} : \; \boldsymbol{\Psi} \mapsto \Psi \in \mathbf{F} \;, \qquad \boldsymbol{\Psi} \in \mathbf{F}_x \;, \tag{3.4}$$

which, in the present context, are unitary maps between each given quantum fibre $\mathbf{F}_x$ and the typical fibre $\mathbf{F}$. For each choice of Poincaré frame $\boldsymbol{u}$ we shall refer to $\Psi$ as the *coordinate wave function* of the local state vector $\boldsymbol{\Psi} \in \mathbf{F}_x$, and to the map in (3.4) as a *soldering map*.

The above definition of the Hilbert bundle $\mathbf{E}$ is totally analogous to that resulting when similar definitions [16,17] of tangent, cotangent, and in general tensor bundles over any differential manifold $\mathbf{M}$ (cf., e.g., [17], Sec. 3.3) are specialized to the case of a Lorentzian manifold $(\mathbf{M}, \boldsymbol{g})$, and the affine frame bundle over $\mathbf{M}$ is contracted to the the Poincaré frame bundle $P\mathbf{M}$ – except that in the latter cases the fibres are finite-dimensional vector spaces. Hence, the bundle $\mathbf{E}$ emerges as a bundle associated with $P\mathbf{M}$, and parallel transport in $\mathbf{E}$ is determined by the choice of affine connection on $P\mathbf{M}$.

From a mathematical point of view, this choice could range over the entire family of affine[2] connections compatible with $\boldsymbol{g}$. However, in a genuine QGR (i.e., quantum general relativity) framework such a connection has to be an affine extension to $P\mathbf{M}$ of the (linear) Levi-Civita connection on the Lorentz frame bundle $L\mathbf{M}$, whose covariant differentiation operators on sections in associated tensor bundles assume the form [18]

$$\nabla_{\boldsymbol{X}} = \partial_{\boldsymbol{X}} + \tfrac{1}{2} \boldsymbol{\omega}_{ij}^{\mathbf{s}}(\boldsymbol{X}) M_{\mathbf{s}}^{ij} \;, \qquad \partial_{\boldsymbol{X}} = X^{\mu} \partial/\partial x^{\mu} \;, \quad \boldsymbol{X} \in T_x \mathbf{M} \;, \tag{3.5}$$

in a vierbein gauge given by a cross-section[3] $\mathbf{s}$ of $L\mathbf{M}$,

$$\mathbf{s} = \{\boldsymbol{e}_i(x) \,|\, i = 0,\ldots,3, \; x \in \mathbf{M}\} \;, \tag{3.6}$$

so that $\boldsymbol{\omega}_{ij}^{\mathbf{s}} = -\boldsymbol{\omega}_{ji}^{\mathbf{s}}$ are the connection 1-forms determined by (cf., e.g., [2], Sec. 2.6)

$$\boldsymbol{\omega}_{ik}^{\mathbf{s}}(\boldsymbol{e}_j) = \tfrac{1}{2}\bigl[\boldsymbol{g}(\boldsymbol{e}_i, [\boldsymbol{e}_j, \boldsymbol{e}_k]) - \boldsymbol{g}(\boldsymbol{e}_j, [\boldsymbol{e}_k, \boldsymbol{e}_i]) + \boldsymbol{g}(\boldsymbol{e}_k, [\boldsymbol{e}_i, \boldsymbol{e}_j])\bigr] \;, \tag{3.7}$$

and $M_{\mathbf{s}}^{ij} = -M_{\mathbf{s}}^{ji}$ are the infinitesimal generators of the Lorentz transformations in the plane spanned by the ($ij$)-axes of the local Lorentz frame $\mathbf{u} = \mathbf{s}(x)$ above the point $x \in \mathbf{M}$ at which the covariant derivative is computed. This must be so since, on one hand, the Levi-Civita connection determines the geodesics in CGR (i.e., classical general relativity) models, and governs the free-fall of classical test particles; whereas, on the other, the behavior of very massive quantum test bodies has to concur in the mean with that of classical test particles. Moreover, since the Levi-Civita connection also determines the parallel transport of local Lorentz frame, by the strong equivalence principle it also determines the inertial (free-falling) moving frames as the frames obtained by the parallel transport of local Lorentz frames along timelike geodesics (cf. [2], Sec. 2.6). Hence, if the strong equivalence principle is to be retained in the QGR regime, then a concept of *quantum* frame has to be introduced, and its infinitesimal parallel transport has to coincide with that in Minkowski space.

---

[2] With regard to linear and affine connections, we follow the terminology introduced in [16].
[3] According to a theorem of Geroch [19], such cross-sections must exist if a spacetime manifold admits a spin structure. For the sake of simplicity, we shall assume this to be the case throughout this paper.

A concept of local quantum frame soldered (cf. [2], Secs. 2.2 and 3.8) to a Poincaré frame $\boldsymbol{u} \in P\mathbf{M}$ above $x \in \mathbf{M}$ is mathematically provided by the family

$$Q_\ell^{\boldsymbol{u}} = \left\{ \boldsymbol{\Phi}_\zeta^{\boldsymbol{u}} \in \mathbf{F}_x \,\Big|\, \zeta = q - iv \in \mathbf{C}^4, \ q \in \mathbf{R}^4, \ v \in \boldsymbol{V}^+ \right\} \qquad (3.8)$$

of generalized coherent states, whose coordinate wave functions are given by

$$\Phi_\zeta(\zeta') = \left( \sigma_x^{\boldsymbol{u}} \boldsymbol{\Phi}_\zeta^{\boldsymbol{u}} \right)(\zeta') = K^{(\ell,m)}(\zeta';\zeta) \ , \qquad (3.9)$$

in terms of the soldering map in (3.4) and the reproducing kernel in (2.12). Due to (2.13) and (2.14), each such family provides a continuous resolution of the identity in the fibre $\mathbf{F}_x$ to which its elements belong, so that

$$\int \left| \boldsymbol{\Phi}_\zeta^{\boldsymbol{u}} \right\rangle d\Sigma(\zeta) \left\langle \boldsymbol{\Phi}_\zeta^{\boldsymbol{u}} \right| = \mathbf{1}_x \ . \qquad (3.10)$$

This means that any vector $\boldsymbol{\Psi}$ from $\mathbf{F}_x$ can be expanded as follows:

$$\boldsymbol{\Psi} = \int \Psi(\zeta) \boldsymbol{\Phi}_\zeta^{\boldsymbol{u}} d\Sigma(\zeta) , \qquad \Psi(\zeta) = \left( \sigma_x^{\boldsymbol{u}} \boldsymbol{\Psi} \right)(\zeta) = \left\langle \boldsymbol{\Phi}_\zeta^{\boldsymbol{u}} \big| \boldsymbol{\Psi} \right\rangle . \qquad (3.11)$$

All the constraints imposed above on a quantum connection are satisfied by the connection $\nabla$ whose operators for covariant differentiation are given by[4]

$$\boldsymbol{\nabla}_{\boldsymbol{X}} = \partial_{\boldsymbol{X}} + i\tilde{\theta}_{\boldsymbol{s}}^i(\boldsymbol{X}) \boldsymbol{P}_{i;\boldsymbol{s}} + \tfrac{i}{2} \tilde{\omega}_{jk}^{\boldsymbol{s}}(\boldsymbol{X}) \boldsymbol{M}_{\boldsymbol{s}}^{jk} \ , \qquad (3.12a)$$

---

[4] Cf. [2], Sec. 4.4. The sign convention in front of the infinitesimal spacetime translations in (3.12a) is dictated by conventions in $M^4$, e.g., whether coordinates are measured in relation to a global Lorentz frame, so that (3.27) results, or in relation to the local Lorentz frames in (3.22), so that the opposite signs are obtained.

$$\partial_X \Psi_x = \int (X^\mu \, \partial \Psi_x(\zeta)/\partial x^\mu) \, \Phi_\zeta^u \, d\Sigma(\zeta) \, , \tag{3.12b}$$

$$\tilde{\theta}_s^i(X) = \theta_s^i(X) + (\nabla_X a)^i \, , \quad \theta_s^i(e_j) = \delta^i{}_j \, , \quad \tilde{\omega}_{jk}^s(X) = \omega_{jk}^s(X) \, , \tag{3.12c}$$

for each choice of Poincaré gauge given by a cross-section

$$s = \{(a(x), e_i(x)) | \, i = 0,\ldots,3, \, x \in \mathbf{M}\} \tag{3.13}$$

of $P\mathbf{M}$, where the covariant derivative in (3.12c) is the one in (3.5) for the vierbein gauge (3.6) with the same tetrads as in (3.13). The infinitesimal generators of spacetime translations and Lorentz transformations that appear in (3.12a) are those corresponding to the following unitary representations,

$$\boldsymbol{U}_{s(x)}(a, \Lambda) = (\sigma_x^s)^{-1} U(a, \Lambda) \sigma_x^s \, , \quad (a, \Lambda) \in \mathrm{ISO}_0(3,1) \, , \tag{3.14}$$

within the fibre $\mathbf{F}_x$. They are explicitly given by the following (unbounded) operators, as they act upon vectors in their domain of definition within that fibre (cf. [2], Sec. 4.4):

$$\boldsymbol{P}_{i;s(x)} = i\partial/\partial q^i \, , \quad \boldsymbol{M}_{s(x)}^{jk} = \boldsymbol{Q}_{s(x)}^j \boldsymbol{P}_{s(x)}^k - \boldsymbol{Q}_{s(x)}^k \boldsymbol{P}_{s(x)}^j \, , \quad \boldsymbol{Q}_{s(x)}^j = q^j - i\partial/\partial m v_j \, . \tag{3.15}$$

As it is the case in any fibre bundle in which a connection is defined, the elements $\Psi_x \in \mathbf{E}$ of a vector field whose domain incorporates a smooth curve $\gamma = \{x(\tau) | \, a \leq \tau \leq b\}$ in $\mathbf{M}$ are said to be parallel transported along $\gamma$ if

$$\nabla_{\dot{x}(\tau)} \Psi_{x(\tau)} = 0 \, , \quad \dot{x}(\tau) = \dot{x}^\mu(\tau) \partial/\partial x^\mu \, , \quad a \leq \tau \leq b \, , \tag{3.16a}$$

or, on account of (3.12a), if equivalently

$$\partial_{\dot{x}(\tau)} \Psi_{x(\tau)} = -i \left( \tilde{\theta}_s^i(\dot{x}(\tau)) \boldsymbol{P}_{i;s} + \tfrac{1}{2} \tilde{\omega}_{jk}^s(\dot{x}(\tau)) \boldsymbol{M}_s^{jk} \right) \Psi_{x(\tau)} \, . \tag{3.16b}$$

If $\gamma$ joins two points $x' \in \mathbf{M}$ and $x'' \in \mathbf{M}$, so that $x' = x(a)$ and $x'' = x(b)$, then this condition determines a unitary operator

$$\tau_\gamma(x'', x') \, : \, \mathbf{F}_{x'} \to \mathbf{F}_{x''} \, . \tag{3.17}$$

As a matter of fact, this operator is completely and unambiguously determined by the parallel transport of quantum frames:

$$\tau_\gamma(x'', x') \, : \, \Psi' = \Psi^\zeta \Phi_\zeta^{u(x')} \, \mapsto \, \Psi'' = \int \Psi(\zeta) \, \tau_\gamma(x'', x') \Phi_\zeta^{u(x')} d\Sigma(\zeta) \, . \tag{3.18}$$

This, in turn, determines in the chosen Poincaré gauge in (3.13) a propagator for parallel transport along $\gamma$,

$$K^s_\gamma(x'',\zeta'';x',\zeta') = \left\langle \Phi^{s(x'')}_{\zeta''} \middle| \tau_\gamma(x'',x')\, \Phi^{s(x')}_{\zeta'} \right\rangle, \tag{3.19}$$

with the following properties, which are mathematically analogous to those in (2.13) and (2.14), respectively:

$$\Psi^\|_{x''}(\zeta'') = \int K^s_\gamma(x'',\zeta'';x',\zeta')\, \Psi_{x'}(\zeta')\, d\Sigma(\zeta'),\qquad \Psi^\|_{x''} = \tau_\gamma(x'',x')\, \Psi_{x'}, \tag{3.20}$$

$$\begin{aligned} K^s_\gamma(x'',\zeta'';x',\zeta') &= \left(K^s_\gamma(x',\zeta';x'',\zeta'')\right)^* \\ &= \int K^s_\gamma(x'',\zeta'';x,\zeta)\, K^s_\gamma(x,\zeta;x',\zeta')\, d\Sigma(\zeta). \end{aligned} \tag{3.21}$$

There is, however, a fundamental distinction between these relations and the corresponding ones in Sec. 2, since the present ones are dependent on the path $\gamma$ in the case where the base manifold $(\mathbf{M},\mathbf{g})$ is curved, and the $q$-integrations are performed over spacelike hypersurfaces within spaces tangent to $\mathbf{M}$, rather than within the manifold $\mathbf{M}$ itself.

On the other hand, if $(\mathbf{M},\mathbf{g})$ coincides with the Minkowski space $M^4$, then parallel transport becomes path-independent, and its outcome coincides with that provided by free quantum propagation.

To see that, let us choose in the Minkowski space $M^4$, viewed as the base manifold of a quantum bundle $\mathbf{E}$, the vierbein gauge represented by the cross-section

$$\mathbf{s}(\mathcal{L})\ :\ x\ \mapsto\ \left\{ \mathbf{e}_i(x)\in T_x M^4 \middle|\ i=0,1,2,3 \right\} \in LM^4,\qquad x = x^i \mathbf{e}_i(O) \in M^4, \tag{3.22}$$

of the Lorentz frame bundle $LM^4$, where each $\{\mathbf{e}_i(x)|\ i=0,1,2,3\}$ is obtained by parallel translating from $O\in M^4$ to $x\in M^4$ all the elements $\mathbf{e}_i(O)$ of a global Lorentz frame $\mathcal{L} = \{\mathbf{e}_i(O)|\ i=0,...,3\}$ with origin at $O$. In the corresponding Poincaré gauge

$$\mathbf{s}(\mathcal{L}) = \left\{ \left(\mathbf{0},\mathbf{e}_i(x)\right)\middle|\ i=0,...,3,\ x\in M^4 \right\} \tag{3.23}$$

we then have

$$\tilde{\theta}^i_{\mathbf{s}(\mathcal{L})} = \theta^i_{\mathbf{s}(\mathcal{L})},\qquad \tilde{\omega}^s_{jk} = 0, \tag{3.24}$$

so that (3.16b) assumes the very simple form

$$\partial_{\dot{x}(\tau)} \Psi_{x(\tau)} = -i\dot{x}^k(\tau)\, P_{k;\mathbf{s}(\mathcal{L})}\, \Psi_{x(\tau)}. \tag{3.25}$$

Consequently, for infinitesimal separations of $x'$ and $x''$, viewed as 4-tuples of Minkowski coordinates in the global Lorentz frame $\mathcal{L}$, we have according to (3.11) and (3.12b) that,

$$\Psi^{\|}_{x''} = \exp\left(-i(x''^j - x'^j)P_j\right)\Psi_{x'}, \qquad P_j = i\partial/\partial q^j. \tag{3.26}$$

This result can be extended immediately to finite separations of $x'$ and $x''$ by segmentations of the path along which the parallel transport is performed. In view of (2.8), this yields

$$\Psi^{\|}_{x''}(\zeta) = \Psi_{x'}(\zeta + x'' - x'), \qquad x', x'' \in \mathbf{R}^4. \tag{3.27}$$

Hence, for the propagator for parallel transport in (3.18) we obtain by using (3.9) and (3.11) that, in the present gauge,

$$K^{s(L)}_{\gamma}(x'', \zeta''; x', \zeta') = \left\langle \sigma^{s(L)}_{x''} \Phi^{s(L;x'')}_{\zeta''} \middle| \sigma^{s(L)}_{x''} \tau_{\gamma}(x'', x') \Phi^{s(L;x')}_{\zeta'} \right\rangle$$

$$= \Phi_{\zeta'}(\zeta'' + x'' - x') = K^{(\ell,m)}(\zeta'' + x''; x' + \zeta'). \tag{3.28}$$

This type of derivation in $M^4$ is even simpler in the Poincaré gauge given by

$$\mathbf{s}_o(L) = \left\{ (\mathbf{a}(x), \mathbf{e}_i(x)) \middle| \mathbf{a}(x) = -x^k \mathbf{e}_k(x), \ x \in M^4 \right\}, \tag{3.29}$$

since, according to (3.12c), in this gauge all the connection coefficients vanish. Hence, in such a gauge the coordinate wave functions do not change under parallel transport, and the same is true of the propagator for parallel transport, with the result that:[5]

$$K^{s_o(L)}_{\gamma}(x'', \zeta''; x', \zeta') = K^{(\ell,m)}(\zeta'' + x'; \zeta' + x') = K^{(\ell,m)}(\zeta''; \zeta'). \tag{3.30}$$

This is understandable, since in this gauge the Poincaré frames in all the spaces tangent to $M^4$ are identifiable with the global Lorentz frame $L$.

## 4. Quantum-geometric boson propagators in curved spacetime

The considerations in the concluding paragraphs of the preceding section indicated that the parallel transport provided by the quantum connection in (3.12) yields, in the special relativistic context, the correct propagators for the free propagation of boson local states. Mathematically, this is due to the fact that parallel transport in Minkowski space is path-independent, and that in the Poincaré gauges (3.29), whose frames can be identified with a single given global Lorentz frame, the propagators for parallel transport are actually independent of the basis points $x \in M^4$; whereas in the vierbein gauges in (3.22), the gauge variables $q$ become redundant after the points they mark in each $T_x M^4$ are identified with the corresponding points in $M^4$. Hence, any averaging of these propagators between two fixed points over various paths leads to the same outcome the propagator along a single path.

---

[5] This formula represents the corrected version of Eq. (6.5) in Chapter 4 of [2].

However, this feature is lost in the presence of an external field (cf. [10], Sec. 2.10). Moreover, from the physical point of view, the parallel transport along a single path cannot provide a physically acceptable formulation of quantum propagation, since, if viewed as a mathematical embodiment of an actual physical process, it would lead to a violation of the uncertainty principle. In fact, the fundamental feature of the "sum-over-path" formulation of quantum propagation obtained by Feynman [5,6] in flat spacetimes by developing an idea of Dirac [20], is that the quantum propagator between two points results from propagation along *all* available paths, provided that particles and antiparticles are simultaneously considered, and that the charge rather than a single particle or antiparticle is being followed.[6] In the absence of a well-defined measure over continuous paths in *real* spacetime (i.e., in the special relativistic context, without passage to the Euclidean regime), these paths are obtained by taking limits of broken polygonal paths [5,6]. The straight line segments of these broken paths have to join sequential pairs of points $x' \in \sigma_{t'}$ and $x'' \in \sigma_{t''}$ on Cauchy surfaces in a time-ordered foliation $\sigma_t$, $-\infty < t < +\infty$, of Minkowski space, and are interpreted as paths of propagation of particles in the case where $\sigma_{t''}$ is in the future in relation to $\sigma_{t'}$, and of antiparticles if the converse is true. Hence, in the case where antiparticles are not included in the considerations, only the first situation occurs.

As mentioned in the Introduction, the adaptation of this fundamental physical idea to the formulation of quantum-geometric propagation in a curved spacetime represented by a globally hyperbolic Lorentzian manifold (**M**,*g*) consists of replacing the straight line segments used in flat spacetimes with the arcs of geodesics of the Levi-Civita connection determined by *g*. Furthermore, since, due to the presence of curvature, the fibres of bundles associated with *L***M** or *P***M** can no longer be identified with their respective typical fibre in a gauge-independent manner, parallel transport has to be employed in the formulation of the transition probabilities between the end-points of the arcs of geodesics constituting the broken paths followed in quantum propagation.

---

[6] Cf. [6], p. 749, where this point is illustrated by Feynman's well-known "bombardier metaphor."

Based on previous studies in curved nonrelativistic Newton-Cartan spacetimes [21,22] as well as in general relativistic spacetimes [23,24], the generic form quantum-geometric propagators for single bosons ultimately adopted in Sec. 4.6 of [2] is obtained as follows.

Let us consider a geometrodynamic evolution [7] that gives rise to a globally hyperbolic spacetime manifold $(\mathbf{M},\mathbf{g})$. Due to the strict determinism inherent in CGR, such an evolution is *mathematically* equivalent to the foliation of $(\mathbf{M},\mathbf{g})$ into a family of Cauchy surfaces $\Sigma_t$, which we shall call reference hypersurfaces. However, we shall view this evolution as an ongoing geometrodynamic process, for which the globally defined ADM parameter $t$ has the *physical* meaning of a natural proper time supplied [2] by the free-fall propagation of the origins of local quantum frames along the geodesics orthogonal to the reference hypersurfaces $\Sigma_t$. Hence, the lapsed "global time" $t'' - t'$ between any two reference hypersurfaces $\Sigma_{t'}$ and $\Sigma_{t''}$ is *locally* determined by the *maximal* proper-time distance between each pair of events lying on those respective hypersurfaces as well as on common timelike geodesics.

Let us consider now any two points $x' \in \Sigma_{t'}$ to $x'' \in \Sigma_{t''}$, and families of hypersurface $\Sigma_{t_n}$, $n = 0,1,...,N$, such that $\varepsilon = t_n - t_{n-1} = (t'' - t')/N \to +0$, where $t_0 = t'$ and $t_N = t''$. We can then construct broken geodesic paths from the geodesic arcs $\gamma(x_{n-1}, x_n)$ of the Levi-Civita connection associated with $\mathbf{g}$, which connect points $x_{n-1} \in \Sigma_{t_{n-1}}$ to points $x_n \in \Sigma_{t_n}$. Then, in a Poincaré gauge given by a cross-section $\mathbf{s} = \{\mathbf{a}(x), \mathbf{e}_i(x) | x \in \mathbf{M}\}$ of $P\mathbf{M}$, the *quantum-geometric propagator* from $x' \in \Sigma_{t'}$ to $x'' \in \Sigma_{t''}$ is defined by the limit

$$\mathbf{K}^{\mathbf{s}}(x'',\zeta'';x',\zeta') = \lim_{\varepsilon \to +0} \int \prod_{n=N}^{1}{}' K^{\mathbf{s}}_{\gamma(x_{n-1},x_n)}(x_n,\hat{\zeta}_n;x_{n-1},\hat{\zeta}_{n-1})\, d\Sigma(x_n,v_n) \, , \quad (4.1a)$$

$$d\Sigma(x_n,v_n) = 2 v_n^\mu\, d\sigma_\mu(x_n) d\Omega(v_n) = 2\mathbf{v}_n \cdot \mathbf{n}_n\, d\sigma(x_n) d\Omega(v_n) \, , \quad (4.1b)$$

$$x_0 = x', \ \hat{\zeta}_0 = \zeta', \ x_N = x'', \ \hat{\zeta}_N = \zeta'', \quad \hat{\zeta}_n = -a(x_n) - i\ell v_n, \ n = 1,2,...,N-1, \quad (4.1c)$$

where the prime on the right-hand side of (4.1a) indicates that $d\Sigma(x_N, v_N)$ is to be omitted from the product, and that therefore no integration is to be performed over $\Sigma_{t''}$.

The measure elements in (4.1b) belong to the invariant phase space measures, which in the Minkowski case reduce to the ones in (2.3). They are constructed from the Riemannian measure over $\Sigma_{t_n}$ with element $d\sigma(x_n)$, the invariant measure over $V^+$ with element $d\Omega(v_n)$, and the normals $\boldsymbol{n}_n$ to $\Sigma_{t_n}$. The 4-tuples $-a(x_n)$ in (4.1c) represent the coordinates of the point of contact between $T_{x_n}\mathbf{M}$ and $\mathbf{M}$ with respect to the Poincaré frame $\{\boldsymbol{a}(x_n),\boldsymbol{e}_i(x_n)\} \in \boldsymbol{s}$.

In the case where $(\mathbf{M},\boldsymbol{g})$ is the Minkowski space $M^4$, this definition reproduces the special relativistic free-fall propagator in (2.12). This is most easily seen in the gauges provided by (3.23) and (3.29).

For example, in the first of these gauges, due to (3.28) we have

$$\boldsymbol{K}^{\boldsymbol{s}(L)}(x'',\zeta'';x',\zeta') = \lim_{\varepsilon \to +0} \int K^{(\ell,m)}(x''+\zeta'';x_{N-1}-i\ell v_{N-1})$$

$$\times \prod_{n=N-1}^{2} d\Sigma(x_n,v_n) K^{(\ell,m)}(x_n - i\ell v_n; x_{n-1} - i\ell v_{n-1}) \, d\Sigma(x_1,v_1) K^{(\ell,m)}(x_1 - i\ell v_1; x'+\zeta') \, , \quad (4.2)$$

so that, on account of (2.14), we obtain

$$\boldsymbol{K}^{\boldsymbol{s}(L)}(x'',\zeta'';x',\zeta') = K^{(\ell,m)}(x''+\zeta'';x'+\zeta') \, , \qquad x',x'' \in \mathbf{R}^4 \, . \qquad (4.3)$$

Hence, according to (3.28), the propagator for parallel transport is recovered in that gauge.

A corresponding result can be obtained for the second of these gauges in the very same manner. However, that result actually immediately follows from the generic Poincaré gauge covariance of the quantum geometric propagator in (4.1).[7]

To establish this generic Poincaré gauge covariance, let us consider for a generic $(\mathbf{M},\boldsymbol{g})$ the transition from any $\boldsymbol{s}$ in (3.13) to any other Poincaré gauge given by some other cross-section $\boldsymbol{s}'$ of of $P\mathbf{M}$. Such a gauge transformation is implemented by the following changes

$$(b(x),\Lambda(x)) \, : \, \boldsymbol{s}(x) \, \mapsto \, \boldsymbol{s}'(x) = \boldsymbol{s}(x) \cdot (b(x),\Lambda(x))^{-1} \, , \qquad (4.4a)$$

$$\boldsymbol{s}(x) = (\boldsymbol{a}(x),\boldsymbol{e}_i(x)) \, , \quad \boldsymbol{s}'(x) = (\boldsymbol{a}'(x),\boldsymbol{e}'_i(x)) \, , \qquad x \in \mathbf{M} \, , \qquad (4.4b)$$

of local Poincaré frames under the action from the right of the Poincaré group. Since the quantum frames are soldered to the Poincaré frames, this Poincaré gauge transformation gives rise to the following change of quantum frames:

---

[7] A different type of approach to these problems was recently presented in [25].

$$\Phi_\zeta^{s(x)} \mapsto \Phi_{\zeta'}^{s'(x)} = \Phi_\zeta^{s(x)} , \qquad \zeta' = b(x) + \Lambda(x)\zeta . \tag{4.5}$$

This is seen to be the case since, by (3.3) and by the definition (3.4) of soldering maps, we have in general that

$$\sigma_x^{u'} = U(b,\Lambda)\sigma_x^u , \qquad u' = u \cdot (b,\Lambda)^{-1} ; \tag{4.6}$$

whereas, according to (3.9),

$$\Phi_{\zeta'}^{u'} = (\sigma_x^{u'})^{-1}\Phi_{\zeta'} = (\sigma_x^u)^{-1} U^{-1}(b,\Lambda)\Phi_{\zeta'} , \tag{4.7}$$

so that, due to (3.9) and the manifest invariance of (2.17a) under Poincaré transformations,

$$\Phi_{\zeta'}(\zeta'') = K^{(\ell,m)}(\Lambda^{-1}(\zeta''-b);\zeta) = (U(b,\Lambda)\Phi_\zeta)(\zeta'') , \qquad \zeta' = b + \Lambda\zeta . \tag{4.8}$$

In turn, (3.19) and (4.5) imply that[8]

$$K_\gamma^{s'}(x'',b(x'') + \Lambda(x'')\zeta'';x',b(x') + \Lambda(x')\zeta') = K_\gamma^s(x'',\zeta'';x',\zeta') . \tag{4.9}$$

This establishes the Poincaré gauge covariance of the propagators for parallel transport, since, in general, the 4-tuples $\zeta$ and $b(x)+\Lambda(x)\zeta$ represent the coordinates of the same point,

$$\zeta = a(x) + \zeta^k e_k(x) = a'(x) + \zeta'^k e_k'(x) \in T_x\mathbf{M}^\mathbf{C} , \qquad \zeta' = b(x) + \Lambda(x)\zeta , \tag{4.10}$$

in the complexified tangent space at $x \in \mathbf{M}$.

This fact, taken in conjunction with the Poincaré gauge invariance of the phase space measure elements in (4.1b), secures the Poincaré gauge covariance of the quantum geometric propagator in (4.1a).

Indeed, in view of (4.9), under the gauge transformation in (4.4) the right-hand side of (4.1a) assumes the form

$$\lim_{\varepsilon \to +0} \int \prod_{n=N}^{1}{}' K_{\gamma(x_{n-1},x_n)}^{s'}(x_n,\tilde{\zeta}_n;x_{n-1},\tilde{\zeta}_{n-1}) \, d\Sigma(x_n,v_n) , \tag{4.11a}$$

$$\tilde{\zeta}_0 = b(x') + \Lambda(x')\zeta', \quad \tilde{\zeta}_N = b(x'') + \Lambda(x'')\zeta'', \quad \tilde{\zeta}_n = -\tilde{a}(x_n) - i\ell\tilde{v}_n , \tag{4.11b}$$

$$-\tilde{a}(x_n) = b(x_n) - \Lambda(x_n)a(x_n) , \quad \tilde{v}_n = \Lambda(x_n)v_n, \quad n = 1,2,...,N-1 , \tag{4.11c}$$

where the 4-tuples in (4.11c) represent the coordinates with respect to the Poincaré frame $\{a'(x_n),e_i'(x_n)\} \in s'$ of, respectively, the point of contact between $T_{x_n}\mathbf{M}$ and $\mathbf{M}$, and the 4-velocity vector $v_n$. However, in view of (4.1b), and of the Lorentz invariance of the measure

---

[8] Cf. [25] for a different method of proving this result, with opposite sign conventions for translations.

over $V^+$ which the phase space measure contains, we have

$$d\Sigma(x_n,v_n) = 2\boldsymbol{v}_n\cdot\boldsymbol{n}_n\, d\sigma(x_n)d\Omega(\tilde{v}_n) = d\Sigma(x_n,\tilde{v}_n) \ , \tag{4.12}$$

with the result that the Poincaré gauge invariance of the quantum-geometric propagator in (4.1a) is thereby established:

$$\boldsymbol{K}^{s'}(x'',b(x'') + \Lambda(x'')\zeta'';x',b(x') + \Lambda(x')\zeta') = \boldsymbol{K}^{s'}(x'',\zeta'';x',\zeta') \ . \tag{4.13}$$

This quantum-geometric propagator obviously satisfies the basic conditions

$$\boldsymbol{K}^s(x'',\zeta'';x',\zeta') = \int \boldsymbol{K}^s(x'',\zeta'';x,\hat{\zeta})\boldsymbol{K}^s(x,\hat{\zeta};x',\zeta')d\Sigma(x,v) \tag{4.14}$$

for quantum propagation. Hence, it can be deemed to govern the propagation of local boson quantum states by providing their coordinate wave functions above various base locations $x$:

$$\Psi_x^s(\zeta) = \int \boldsymbol{K}^s(x,\zeta;x_0,\zeta')\, \Psi_{x_0}^s(\zeta')d\Sigma(\zeta') \ , \qquad x \in \Sigma_t \ . \tag{4.15}$$

Extrapolating from (2.9), we can interpret

$$P(\boldsymbol{\Psi}_{x_0};B) = \int_B \left|\Psi_x^s(-a(x) - i\ell v)\right|^2 d\Sigma(x,v) \ , \qquad B \subset \Sigma_t \ , \tag{4.16}$$

as the *conditional* probability that, if a boson in the *local* quantum state $\Psi_{x_0}\in \mathbf{F}_{x_0}$ were prepared at $x_0 \in \Sigma_{t_0}$, then it would be later detected within the (Borel) set $B$ along $\Sigma_t$.[9] However, whereas (2.9) provides an absolute probability, which is conserved on account of the fact that the special relativistic quantum propagation is governed by a group of unitary operators in a *single* Hilbert space, in the case of a curved base manifold $(\mathbf{M},\boldsymbol{g})$ the probabilities in (4.16) are only relative, since then there is no Poincaré gauge invariant identification of all fibres $\mathbf{F}_x$ with the typical fibre $\mathbf{F}$, and consequently there is no single generator of quantum evolution that plays the role of Hamiltonian. Physically, this resolves (cf. [2], Sec. 4.7) the difficulties with conventional treatment of quantum propagation in curved spacetime, which in non-static spacetimes lead to *ex nihilo* particle creation even in free-fall situations (cf. [26], Sec. 3.3).

We note that, in accordance with Feynman's ideas [6], the broken paths of quantum propagation incorporated in (4.1a) are not restricted to the timelike ones. This means that the Einstein causality that we observe macroscopically in weak gravitational fields is due to the superposition of probability amplitudes at points in classically forbidden regions, resulting in mutual cancellations within those regions. In strong gravitational fields, such as those of black holes, this is might no longer be the case, as suggested by the physical interpretation [27] of black hole evaporation.

---

[9] The generalization to measurements involving also the stochastic 4-velocities $\boldsymbol{v}$ at various locations in $\mathbf{M}$ is very straightforward (cf. [2], Sec. 4.7).

This makes it clear that in the generic case propagation along spacelike paths cannot be neglected. Hence, in general the computation of the quantum geometric propagator in (4.1a) requires the knowledge of propagators for parallel transport along arbitrary paths.

## 5. Quantum-geometric propagation for Robertson-Walker cosmologies

In general, the wave functions for parallel transported local states can be obtained by solving within the typical fibre **F** the counterpart of the equation (3.16b),

$$d\Psi_{x(\tau)}/d\tau = \dot{x}^\mu(\tau)\partial\Psi_x/\partial x^\mu = -iD_s(\tau)\Psi_{x(\tau)}, \tag{5.1a}$$

$$D_s(\tau) = \tilde{\theta}_s^i(\dot{x}(\tau))P_i + \tfrac{1}{2}\tilde{\omega}_{jk}^s(\dot{x}(\tau))M^{jk} . \tag{5.1b}$$

which is obtained by applying the soldering map (3.4) to both sides of that equation. The infinitesimal generators in (5.1b) are, in accordance with (3.14), the unitary transforms under the corresponding soldering maps of those in (3.15), so that they are given by

$$P_i = i\partial/\partial q^i, \quad M^{jk} = Q^j P^k - Q^k P^j, \quad Q^j = q^j - i\partial/\partial m v_j, \tag{5.1c}$$

as they act on coordinate wave functions in the typical fibre **F**. Hence, the operator in (5.1b) is (essentially) self-adjoint. We therefore recognize in (5.1a) the equivalent of a time-dependent Schrödinger equation, whose general solution along any given smooth curve $\gamma = \{x(\tau)|\, a \leq \tau \leq b\}$ with end-points $x' = x(a)$ and $x'' = x(b)$ is

$$\Psi_{x''}^{\|} = T\exp\left[-i\int_a^b D_s(\tau)d\tau\right]\Psi_{x'} . \tag{5.2}$$

One way to compute this solution is to subdivide $\gamma$ into segments corresponding to the values $a = \tau_0 < \tau_1 < \cdots < \tau_N = b$, of the generic parameter $\tau$, and then write

$$\Psi_{x''}^{\|} = \lim_{\varepsilon \to +0}\prod_{n=N}^{1}\exp[-i\varepsilon D_s(\tau_n)]\Psi_{x'} , \quad \varepsilon = \tau_n - \tau_{n-1}, \tag{5.3a}$$

taking note of the fact that, to the first order in $\varepsilon$,

$$\Psi^{\|}_{x_n}(\zeta) = \left(\exp[-i\varepsilon D_s(\tau_n)]\Psi^{\|}_{x_{n-1}}\right)(\zeta) = \Psi^{\|}_{x_{n-1}}(\varepsilon\dot{x}(\tau_n) + \Lambda_n(\varepsilon)\zeta) , \qquad (5.3b)$$

$$\Lambda_n(\varepsilon) = \exp\left(\tfrac{1}{2}\varepsilon\omega^s_{jk}(\dot{x}(\tau_n))M^{jk}\right) \in SO_0(3,1) , \qquad (5.3c)$$

where the $4 \times 4$ matrices in (5.3c) are the generators of infinitesimal Lorentz transformations in the ($jk$)-plane, and represent the counterparts in the typical fibre of $T\mathbf{M}^C$ of those in (3.5).

As a relatively simple but interesting illustration of the resulting method for computing the quantum geometric propagators in (4.1), let us consider the case of such propagators for a base Lorentzian manifold $(\mathbf{M},\mathbf{g})$ with a spatially flat Robertson-Walker metric [7],

$$\mathbf{g} = dt \otimes dt - a^2(t)\delta_{\alpha\beta}dx^\alpha \otimes dx^\beta , \qquad \alpha,\beta = 1,2,3 , \qquad (5.4)$$

where $a(t)$ is an expansion (or contraction) factor depending on a cosmological time $t$ which as mentioned in Sec. 4, is determined locally. It is then natural to consider in the resulting Lorentz bundle $L\mathbf{M}$ the following vierbein and its dual

$$\mathbf{e}_0 = \partial/\partial t , \quad \mathbf{e}_\alpha = a^{-1}(t)\partial/\partial x^\alpha , \qquad (5.5a)$$

$$\boldsymbol{\theta}^0 = dt , \qquad \boldsymbol{\theta}^\alpha = a(t)dx^\alpha . \qquad (5.5b)$$

In the sequel we shall work exclusively with the Poincaré gauge

$$\mathbf{s} = \left\{(\mathbf{0},\mathbf{e}_i(x)) \mid i = 0,...,3, \ x \in \mathbf{M} \subset \mathbf{R}^4\right\} \qquad (5.5c)$$

determined by the above vierbein, so that we shall omit the $\mathbf{s}$-superscripts which earlier indicated the gauge dependence of connection 1-forms and other previously introduced geometric objects. We shall refer to the above Poincaré gauge as the standard gauge of such a cosmological model.

The Lie brackets of the tetrads in this standard gauge obviously are:

$$[\mathbf{e}_0,\mathbf{e}_\alpha] = -(\dot{a}/a)\mathbf{e}_\alpha , \quad [\mathbf{e}_\alpha,\mathbf{e}_\beta] = 0, \qquad \dot{a} = da/dt . \qquad (5.6)$$

Hence, from (3.7) we can easily compute that

$$\boldsymbol{\omega}_{0\alpha}(\mathbf{e}_0) = 0 , \quad \boldsymbol{\omega}_{0\alpha}(\mathbf{e}_\beta) = (\dot{a}/a)\delta_{\alpha\beta} , \quad \boldsymbol{\omega}_{\alpha\beta} \equiv 0 , \qquad (5.7)$$

so that we have in general

$$\tilde{\theta}^i(X) = \theta^i(X) = X^i , \qquad X = X^i\mathbf{e}_i \in T\mathbf{M} , \qquad (5.8a)$$

$$\tilde{\omega}_{0\alpha}(X) = (\dot{a}/a)X^\alpha \,, \qquad \tilde{\omega}_{\alpha\beta}(X) = 0 \,, \qquad \alpha,\beta = 1,2,3 \,. \tag{5.8b}$$

A feature common to all Robertson-Walker models is that for all their geodesics

$$\gamma = \left\{ (t(\tau),\mathbf{x}(\tau)) \big| \tau \in [a,+\infty) \subset \mathbf{R}^1 \right\} \,, \qquad \mathbf{x} = (x^1,x^2,x^3) \,, \tag{5.9}$$

$(t,\mathbf{x}(\tau))$ traces in each reference hypersurface $\Sigma_t$ a geodesic with respect to the metric induced there by $\boldsymbol{g}$ [28], which in the present spatially flat case is

$$\boldsymbol{g}_t = a^2(t)\delta_{\alpha\beta}\, dx^\alpha \otimes dx^\beta \,. \tag{5.10}$$

As a consequence of the geodesic equations, we have [28]

$$\frac{d^2 t}{d\tau^2} + a(t)\dot{a}(t)\,\boldsymbol{g}_t\!\left(\frac{d\mathbf{x}}{d\tau},\frac{d\mathbf{x}}{d\tau}\right) = 0 \,, \tag{5.11a}$$

$$\frac{d^2\mathbf{x}}{d\tau^2} + 2\frac{\dot{a}(t)}{a(t)}\frac{dt}{d\tau}\frac{d\mathbf{x}}{d\tau} = 0 \,. \tag{5.11b}$$

For spatially flat Robertson-Walker models the geodesics in each $\Sigma_t$ are obviously straight lines, so that we can set in (5.11b) for some fixed initial-data hypersurface $\Sigma_{t_0}$,

$$\mathbf{x}(\tau) = \mathbf{A}r(\tau) + \mathbf{B} \,, \qquad \frac{d\mathbf{x}}{d\tau} = \mathbf{A}\frac{dr}{d\tau}\,, \qquad \frac{d^2\mathbf{x}}{d\tau^2} = \mathbf{A}\frac{d^2 r}{d\tau^2}\,, \tag{5.12}$$

and thus obtain a simple system of two nonlinear equations for $t(\tau)$ and $r(\tau)$.[10]

Due to the spatial rotational symmetry of the present Robertson-Walker models, the propagators for parallel transport along arbitrary geodesic arcs $\gamma(x_{n-1},x_n)$ in (4.1a) can be obtained by space rotations from those for geodesics corresponding to constant values of $x^2$ and $x^3$. Furthermore, for the computation of (4.1a) it is desirable to adopt $t$ instead of the affine parameter $\tau$ as a new parameter for each $\gamma(x_{n-1},x_n)$ – with these two parameters coinciding only for the geodesics that are orthogonal to the reference hypersurfaces.

Let us, therefore, consider any geodesic arc $\gamma = \{x(t)|\, a \le t \le b\}$ with end-points $x'$ = $x(a)$ and $x'' = x(b)$, for which $x^0(t) = t$, and for which the coordinates $x^2$ as well as $x^3$ are constant. Along such a geodesic arc we shall have

$$\tilde{\theta}^0(\dot{x}(t)) = 1, \qquad \tilde{\theta}^1(\dot{x}(t)) = \dot{x}^1(t) \,, \qquad \tilde{\omega}_{01}(\dot{x}(t)) = \dot{a}(t)\dot{x}^1(t)/a(t), \tag{5.13}$$

---

[10] In the spatially flat case, equally simple is to solve directly the geodesic equations for fixed $x^2$ and $x^3$. Solutions for the generic Robertson-Walker timelike case can be found in [29].

with the values assumed by the remaining connection coefficients in (5.1b) being equal to zero. Consequently, by (5.3) we have, to the first order in $\varepsilon$,

$$\Psi^{\|}_{x_n}(\zeta) = \Psi^{\|}_{x_{n-1}}(\varepsilon \dot{x}(t_n) + \exp[\varepsilon(\dot{a}(t_n)/a(t_n))\dot{x}^1(t_n)M^{01}]\zeta) , \tag{5.14}$$

where $M^{01}$ is the generator of Lorentz boosts in the $q^1$-direction, and therefore it is represented by a $4 \times 4$ matrix whose (01) and (10) elements are equal to one, and all the remaining elements are equal to zero. Using induction, and then taking the limit $\varepsilon \to +0$ and $N \to +\infty$, we obtain that

$$\Psi^{\|}_{x''}(\zeta) = \Psi_{x'}(c_\gamma + \Lambda_\gamma q - i\ell\Lambda_\gamma v) , \tag{5.15a}$$

where the Poincaré transformation on the right-hand side of (5.15a) is given by

$$c_\gamma = \int_a^b \exp[M^{01}\int_t^b \dot{a}(s)\dot{x}^1(s)/a(s)ds]\dot{x}(t)dt , \tag{5.15b}$$

$$\Lambda_\gamma = \exp[M^{01}\int_a^b \dot{a}(t)\dot{x}^1(t)/a(t)dt] . \tag{5.15c}$$

The physical interpretation of this result is very clear: due to the geometrodynamic expansion (or contraction) in the base manifold, any of the local Lorentz frames of the standard gauge in (5.5) when in free fall along a geodesic undergoes a Lorentz boost given by the inverse of $\Lambda_\gamma$ in relation to the local frames of that gauge; the quantum frames soldered to that moving frame undergoes an additional internal spacetime translation in the amount $-c_\gamma$. In case that $a(t) \equiv 1$, so that one is dealing with Minkowski spacetime, there is no Lorentz boost, and the translation is in the amount $x' - x'' \in \mathbf{R}^4$, so that (3.27) is recovered.

It might appear that when the above result is used along the geodesic arcs $\gamma(x_{n-1}, x_n)$ in (4.1a), only the contributions from $x_n$ in the causal future of $x_{n-1}$ lead to significant superposition effects, so that when the limit $\varepsilon \to +0$ is taken in (4.1a) only the infintesimal form of (5.15) would be required. However, due to the presence of the fundamental length $\ell$, fluctuations outside that causal future manifest themselves, since the prepared state is not perfectly sharply localized. In the Minkowski case corresponding to $a(t) \equiv 1$, such fluctuations are negligible beyond several orders of magnitude of $\ell$ [30]. On the other hand, as we already mentioned in Sec. 4, in strong gravitational fields the situation can be quite different due to "tunnel effects." Therefore, it cannot be taken for granted that a large rate of expansion or contraction might not also lead to some type of quantum violations of strict Einstein causality. Hence, this remains an interesting open problem.

**Acknowledgements.**

The author would like to thank Professor W. Drechsler and Dr. P. A. Tuckey for sending him pre-publication versions of reference [25], and for the ensuing correspondence, which has primarily motivated the writing of the present paper.